\documentclass[aps,prb,singlecolumn,showpacs,superscriptaddress]{revtex4-1}
\usepackage[latin1]{inputenc}
\usepackage{amsmath}
\usepackage{amssymb}
\usepackage{siunitx}
\usepackage[english]{babel}
\usepackage{graphicx, color}
\usepackage{hyperref}
\definecolor{link}{rgb}{0.1,0.1,0.9}
\hypersetup{colorlinks=true,linkcolor=link,citecolor=link,urlcolor=link,linktocpage}
\usepackage{epstopdf}
\usepackage{color}
\usepackage{longtable}
\usepackage{multirow}

\begin{document}
	
\title{\textbf{A magnetocaloric study on the series of 3d-metal chromites ACr$_{2}$O$_{4}$ where A = Mn, Fe, Co, Ni, Cu and Zn}}
	
\author{Anzar Ali}
\email{anzarali@iisermohali.ac.in}
\affiliation{Department of Physical Sciences, Indian Institute of Science Education and Research, Knowledge city, Sector 81, SAS Nagar, Manauli PO 140306, Mohali, Punjab, India}

\author{Yogesh Singh} 
\affiliation{Department of Physical Sciences, Indian Institute of Science Education and Research, Knowledge city, Sector 81, SAS Nagar, Manauli PO 140306, Mohali, Punjab, India}

\begin{abstract}
The 3d-metal chromites ACr$_2$O$_4$ where A is magnetic ion, show the paramagnetic to ferrimagnetic phase transition at T$_C$ while for non magnetic A-site ion, ACr$_2$O$_4$ show paramagnetic to antiferromagnetic phase transition at T$_N$. In this report, we present the detailed study of magnetic and the magnetocaloric effect (MCE) of the 3d-metal chromites ACr$_{2}$O$_{4}$ (where A = Mn, Fe, Co, Ni, Cu and Zn) near  T$_C$ and T$_N$. We find the magnitude of MCE (-$\Delta$S$_M$) decreases on decreasing the magnetic moment of A-site ion with a exception for CuCr$_{2}$O$_{4}$ . Additionally, to know more about the order and nature of phase transition, we have made a scaling analysis of (-$\Delta$S$_{M}$) for all the chromites across the phase transition temperatures T$_C$ and T$_N$.
 
\end{abstract}

\maketitle

\section{Introduction}

Spinel compounds with general formula AB$_{2}$O$_{4}$ \cite{Bra}, where A-site is occupied by divalent cations and the B-site is occupied with trivalent cations, have attracted much attention in recent years due to their large magnetocapacitance \cite{Hem, Web}, colossal magnetoresistance \cite{Ram, Web, Leh} and tunable magnetocaloric effect \cite{Fra, Dey_1, Ali}. In spinel chromites ACr$_{2}$O$_{4}$ (A= 3d transitional metals, Mn, Fe, Co, Ni, Cu and Zn), the spin, orbital, and lattice degree of freedom play an essential role in   the enhancement of  multifunctional behavior such as magnetoelastic \cite{Lee, Lap}, magnetodielectric \cite{Muf}, multiferroic \cite{Eer, Bib, Che, Ram_2, Fie} and magnetocaloric effect \cite{Fra}. A deeper understanding of the interactions between spin, lattice, and orbital may provide a great way to use these spinels chromites with their fullest application potential.

The spinel chromites ACr$_{2}$O$_{4}$ where A-site is non-magnetic (Zn, Mg, Cd) show a high degree of frustration \cite{Moe}. In such compounds, the antiferromagnetic nearest neighbor interaction between Cr$^{3+}$ ions do not order to the lowest measured temperatures \cite{Cher}. The spinel chromites  where A-site is a magnetic 3d-transition metal ion show a ferrimagnetic ordering on cooling below some specific temperatures.The coupling between spin, lattice, and orbital degree of freedom leads to several magnetostructural transitions. All the spinels chromites ACr$_{2}$O$_{4}$ where A-site is Mn, Fe, Co, Ni, and Cu go through the ferrimagnetic ordering at the temperature T$_C$ = 41 \cite{Dey_2}, 101 \cite{Mai}, 97 \cite{Tom_1}, 68 \cite{Tom_2}, and 129K \cite{Gur} respectively, while ZnCr$_{2}$O$_{4}$ order antiferromagnetically at T$_N$ = 13K \cite{You}. The coupled magnetostructural transitions have been observed and studied well for the chromites such as MnCr$_{2}$O$_{4}$ \cite{Muf}, FeCr$_{2}$O$_{4}$ \cite{Kir}, CoCr$_{2}$O$_{4}$ \cite{Muf}, and NiCr$_{2}$O$_{4}$ \cite{Tay}. In such spinels chromites with magnetic A-site, the structural changes occur at the ferrimagnetic ordering where system goes from tetragonal to orthorhombic symmetry \cite{Mat,Stu}. In general, the magnetostructural transitions are of first order in nature. In first order transitions, the order parameter changes abruptly and hence a massive change in the entropy of a system at the transition point is expected.

Generally, the rare-earth-based compounds have a large effective magnetic moment and show a giant magnetocaloric effect \cite{PhysRevLett.78.4494, PhysRevB.72.174420, C5DT01254F}. A lot of experimental, theoretical work has been carried out to understand the magnetocaloric behavior of rare-earth-based materials \cite{doi:10.1002/adma.200901072, Ali_2019}. The rare-earth-based based materials are expensive and get sometimes get oxidized in air. So it is customary to search materials which cost less and have higher stability in air. Recently, the transition metal oxide compounds have attracted much attention due to their interesting multifunctional behavior such as multiferroic and magnetocaloric effect. These interesting physical properties are due to an interplay among different degrees of freedom, such as spin, lattice, and orbital. The study of MCE on the series of spinel chromites may give some indications to understand and control such multifunctional behavior.

The work we are presenting in this report includes the comprehensive study of the magnetocaloric effect in the vicinity of the ferrimagnetic phase transition for ACr$_{2}$O$_{4}$ (A = Mn, Fe, Co, Ni, and Cu) and across the antiferromagnetic transition for ZnCr$_{2}$O$_{4}$.  We also include MCE results on NiCr$_{2}$O$_{4}$ from the previous reported work \cite{Ali} to be able to make a comparison. Additionally we have also made a entropy scaling analysis across these transitions to get insight into the governing mechanism and order of the magnetic phase transition.          

\section{Experimental Details}

The polycrystalline samples of spinels ACr$_{2}$O$_{4}$ where A = Mn, Fe, Co, Ni, Cu and Zn have been synthesized using solid state reaction method.  The starting materials AO (99:995\%, Alfa Aesar) and Cr$_{2}$O$_{3}$ (99:999\%, Alfa Aesar) were taken in the stoichiometry ratio. The reactants were mixed thoroughly and then pelletized using 5-ton pressure. All the pelletized materials then placed in alumina crucibles covered with a lid except MnCr$_{2}$O$_{4}$ and FeCr$_{2}$O$_{4}$. The pelletized mixture of MnCr$_{2}$O$_{4}$ and FeCr$_{2}$O$_{4}$ were sealed in the evacuated quartz tube and then loaded into box furnace. Several heat treatments with intermediate grindings were given with temperatures as mentioned in the Table \ref{tab_spinel}.

Room temperature powder X-ray diffraction of all spinels ACr$_{2}$O$_{4}$ where A = Mn, Fe, Co, Ni, Cu and Zn compounds are refined by Rietveld refinement method using GSAS software. The spinels  MnCr$_{2}$O$_{4}$,  FeCr$_{2}$O$_{4}$,  CoCr$_{2}$O$_{4}$, and  ZnCr$_{2}$O$_{4}$ crystallize in cubic group Fd$\bar{3}$m while NiCr$_{2}$O$_{4}$ and  CuCr$_{2}$O$_{4}$, which contain Jahn-Teller active Ni$^{2+}$ and Cu$^{2+}$ ions crystallize in tetragonal space group I4$_{1}$/amd. The refined lattice parameters are listed in Table (\ref{6_XRD}) and match well with the literature \cite{Dey_1, Kir, Tay, Gur, Lee_2}.The magnetic measurements were performed using Quantum Design (QD) physical property measurement system (PPMS).

\section{Result and Discussion}

\begin{figure*}[htbp]
	\centering
	\includegraphics[width=\linewidth]{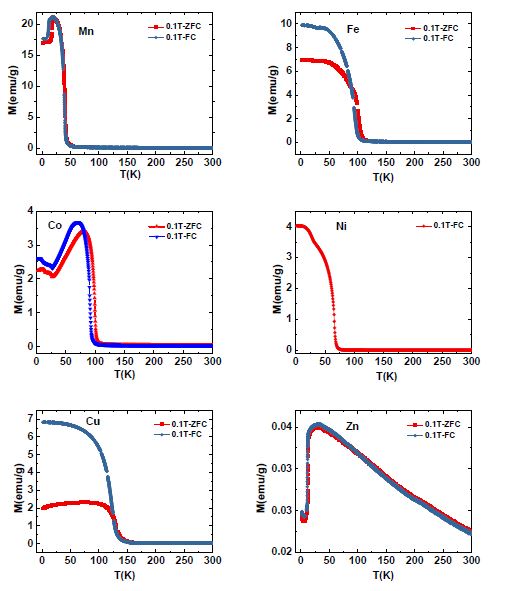}
	\caption{ZFC and FC magnetization curves recorded under magnetic field H = 0.1T for all the compounds indicated in the plots.}
	\label{fig: MT}
\end{figure*}

Figure \ref{fig: MT} shows magnetization versus temperature, M(T), measured under zero-field cooled (ZFC) and field cooled (FC) protocol in an applied field of 0.1T. The sharp increase of magnetization at temperature T$_C$ = 41, 101, 97, 68, and 129K in MnCr$_{2}$O$_{4}$,  FeCr$_{2}$O$_{4}$,  CoCr$_{2}$O$_{4}$, NiCr$_{2}$O$_{4}$, and CuCr$_{2}$O$_{4}$ respectively indicate the ferromagnetic or ferrimagnetic ordering. These transitions are well studied in literature and identified as ferrimagnetic ordering between A$^{2+}$ and Cr$^{3+}$. ZnCr$_{2}$O$_{4}$ order antiferromagnetically at T$_N$ = 13K.

\begin{figure*}[htbp]
	\centering
	\includegraphics[width=\linewidth]{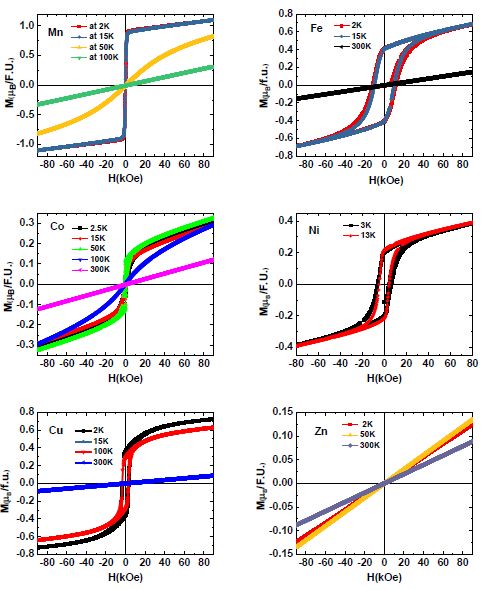}
	\caption{Isothermal curves of magnetization verses magnetic field for all the ACr$_{2}$O$_{4}$ where A = Mn, Fe, Co, Ni, Cu and Zn compounds at various temperatures indicated in the plots.}
	\label{fig: MH_1}
\end{figure*}

Figure \ref{fig: MH_1} shows magnetization versus field, M(H) measured at various temperatures indicated in the plots for all the spinels MnCr$_{2}$O$_{4}$,  FeCr$_{2}$O$_{4}$,  CoCr$_{2}$O$_{4}$, NiCr$_{2}$O$_{4}$, CuCr$_{2}$O$_{4}$, and ZnCr$_{2}$O$_{4}$ compounds. All the spinels shows the characteristic of soft ferrimagnet with a tendency for rapid saturation accompanied by a small coercivity except ZnCr$_{2}$O$_{4}$, which shows a linear field dependent magnetization, which is typical behavior for an antiferromagnetic phase. In these spinels FeCr$_{2}$O$_{4}$ shows highest coercivity and MnCr$_{2}$O$_{4}$ shows minimum coercivity. The M(H) however, never saturates even at our highest applied magnetic fields suggesting that there is a paramagnetic or antiferromagnetic component at all measured temperatures. This suggests that some magnetic ions do not participate in ferrimagnetic ordering.

\begin{figure*}[htbp]
	\centering
	\includegraphics[width=\linewidth]{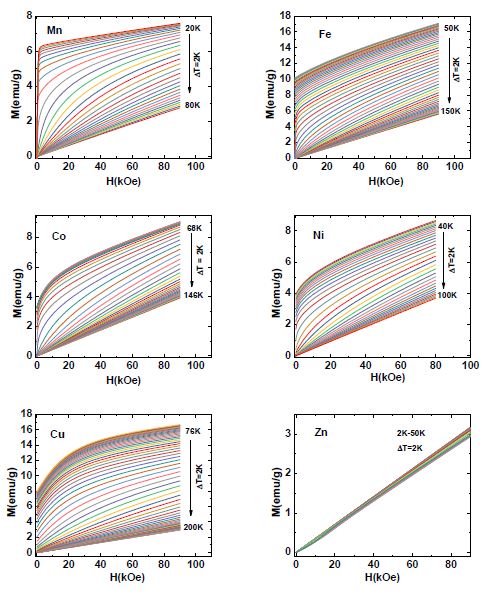}
	\caption{Series of isotherms of magnetization with a step size of $\Delta T = 2K$  for ACr$_{2}$O$_{4}$ where A = Mn, Fe, Co, Ni, Cu and Zn compounds in the temperature ranges indicated in the plots.}
	\label{fig: MH_2}
\end{figure*}

To understand field dependent magnetic behavior of these samples over a broad temperature range and to calculate its magnetocaloric potential, M(T, H) is obtained from measurement of M vs H at various temperatures. Figure \ref{fig: MH_2} shows the series of isotherms M(T,H) measured for all the spinel compounds in the temperature and field ranges indicated in the plots.  M(T,H) isotherms were measured at different temperature with 2 K interval. When an external magnetic field is applied to a magnetic material, its atom's magnetic moment try to align along the applied magnetic field and hence its magnetic entropy decreases . Under adiabatic process , the temperature of the material increases. Conversely when applied magnetic filed is removed adiabatically, the temperature of material decreases. This thermal response of magnetic material in the varying magnetic field is called magnetocaloric effect (MCE).  The magnetocaloric effect is related to the change in magnetic entropy $-\Delta S_{M}$ and it can be calculated by applying Maxwell's thermodynamic relation as stated below \cite{Fra}:

\begin{equation}
\Delta S_M (T, H_{0 \rightarrow H_{MAX}}) = \mu_{0}\int _{0}^{H_{MAX}} \left| \dfrac{dM}{dT}\right|_H dH
\label{maxwell}
\end{equation}

Another important parameter of a MCE material is the cooling efficiency also called relative cooling power (RCP) \cite{Woo}. RCP is defined by the expression:

\begin{equation}
RCP = -\Delta S_{M}(T,H) \times \Delta T_{FWHM}
\label{RCP}
\end{equation}

\begin{figure*}[htbp]
	\centering
	\includegraphics[width=\linewidth]{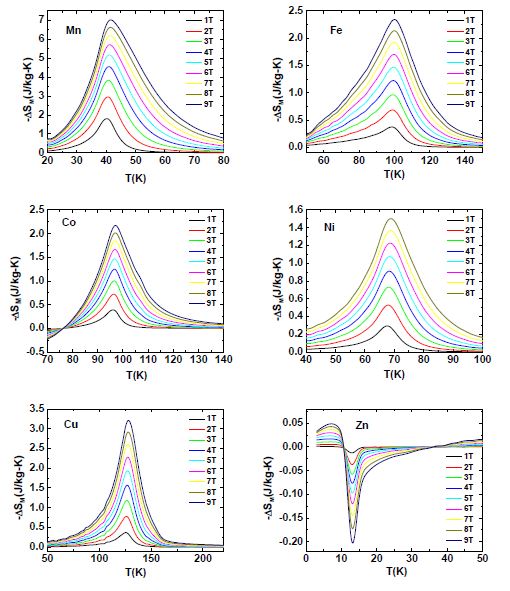}
	\caption{The thermal profile of field induced change in magnetic entropy, $-\Delta S_{M}$, calculated from Maxwell's equation \ref{maxwell} using isothermal magnetization curves in the vicinity of transition temperature for all the ACr$_{2}$O$_{4}$ where A = Mn, Fe, Co, Ni, Cu and Zn compounds  at various external magnetic field as indicated in the plots.}
	\label{fig: MCE}
	
\end{figure*}

The temperature dependence of $-\Delta S_{M}$  vs T thus obtained from a series of isothermal magnetization at various temperature for all the spinels across the magnetic transition at T$_{C}$ and T$_{N}$ are shown in figure \ref{fig: MCE}. The magnitude of the peak of  $-\Delta S_{M}$ increases with increasing magnetic field as expected. The temperature range for significant magnetocaloric effect can be estimated from the full width at half maximum (FWHM) of  $-\Delta S_{M}$. The values of maximum entropy change $-\Delta S_{M}$, $\Delta T_{FWHM}$, RCP and transition temperature ($T_{C}$) have been calculated for all the spinel samples. All the ACr$_{2}$O$_{4}$ where A is magnetic ion show the positive value of maximum entropy change (-$\Delta S_{M}$) while ZnCr$_{2}$O$_{4}$ has negative value of maximum entropy change (-$\Delta S_{M}$) which is expected across AFM-PM transition. The magnetocaloric parameters for all the ACr$_{2}$O$_{4}$ where A = Mn, Fe, Co, Ni, Cu and Zn compounds in a external magnetic field of strength 9T, together with some other spinels materials are listed in the Table \ref{tab} for comparison. The maximum entropy change (-$\Delta$S$_M$) and RCP increase with the increase of A-site spin magnetic moment and we get highest -$\Delta$S$_M$ and RCP for MnCr$_2$O$_4$. This behavior suggest that the degree of magnetization in these spinels near paramagnetic to ferrimagnetic phase transition is highly coupled with the A-site magnetic moment. However, this trend deviate for the spinel CuCr$_2$O$_4$. To find the origin of this deviation further microscopic investigation are needed.  

\begin{figure*}[htbp]
	\centering
	\includegraphics[width=\linewidth]{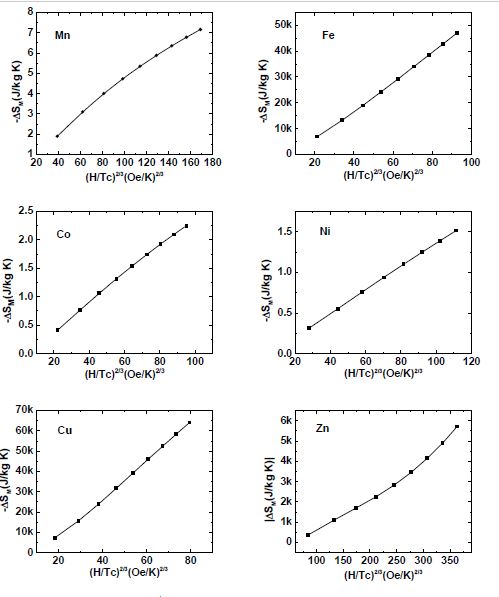}
	\caption{ $\Delta S_M^{max}$ versus $H^{2/3}$ data showing a linear dependence expected for a mean-field transition.}
	\label{PL_1}
	
\end{figure*}

In order to get insight about the nature of the magnetic transition in the vicinity of the transition temperature we have done scaling analysis of the maximum entropy change. According to mean field model the magnetic entropy change at the transition is expected to follow a power law behavior given by |$\Delta S_M|\propto H^n$ with $n = 2/3$~ \cite{Oes}.  Figure \ref{PL_1} shows the plot of $\Delta S_M$ versus $H^{2/3}$ for the magnetic transition at $T_{0}$ for all the  spinels  MnCr$_{2}$O$_{4}$,  FeCr$_{2}$O$_{4}$,  CoCr$_{2}$O$_{4}$, NiCr$_{2}$O$_{4}$, CuCr$_{2}$O$_{4}$, and ZnCr$_{2}$O$_{4}$ compounds. FeCr$_{2}$O$_{4}$,  CoCr$_{2}$O$_{4}$, and NiCr$_{2}$O$_{4}$ show a nearly linear plot, which strongly suggests that these transition are mean-field like. The plots $\Delta S_M$ versus $H^{2/3}$ for  MnCr$_{2}$O$_{4}$,   CuCr$_{2}$O$_{4}$, and  ZnCr$_{2}$O$_{4}$ deviate from linearity, indicating non mean filed like transitions in these compounds.

Franco \textit{et al.} \cite{Fra_3} gave a model for the materials which has second order magnetic phase transition. They proposed that the $\Delta S_M(T)$ curves at different magnetic fields are expected to collapse onto a common universal curve when they are plotted as $\Delta S_M/\Delta S_M^{max}$ versus $\theta$, where $\Delta S_M^{max}$ is the value of $\Delta S_M$ at the transition temperature around which the scaling analysis is being made, and $\theta$ is a reduced temperature given by $\theta = -{T - T_c \over T_{r1} - T_c}$ for $T \leq T_c$ and $\theta = {T - T_c \over T_{r2} - T_c}$ for $T > T_c$.  The $T_{r1}$ and $T_{r2}$ are the temperatures at the full width at half maximum of the anomaly in $\Delta S_M$~\cite{2}.  We have performed the above scaling analysis for all the spinels MnCr$_{2}$O$_{4}$,  FeCr$_{2}$O$_{4}$,  CoCr$_{2}$O$_{4}$, NiCr$_{2}$O$_{4}$, CuCr$_{2}$O$_{4}$, and ZnCr$_{2}$O$_{4}$ compounds.  

\begin{figure*}[htbp]
	\centering
	\includegraphics[width=\linewidth]{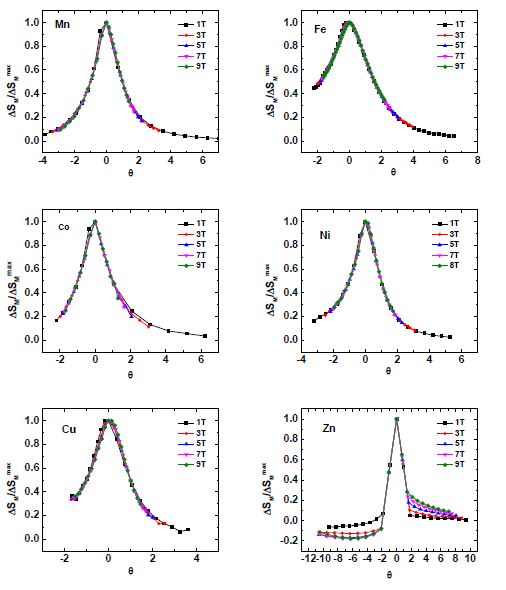}
	\caption{The collapse of all the $\Delta S_M$ data at different magnetic fields onto a universal curve when plotted as $\Delta S_M/\Delta S_M^{max}$ vs $\theta$}
	\label{PL_2}
	
\end{figure*}

The plot of $\Delta S_M/\Delta S_M^{max}$ at various fields versus the reduced parameter $\theta$ is shown in Fig.~\ref{PL_2} for all the compounds.  We see that all the $\Delta S_M$ curves for the different magnetic fields approximately collapse onto a single universal master curve except for ZnCr$_{2}$O$_{4}$ .  This is strong evidence of the second order nature of the magnetic transition at $T_{C}$ in spinels MnCr$_{2}$O$_{4}$,  FeCr$_{2}$O$_{4}$,  CoCr$_{2}$O$_{4}$, NiCr$_{2}$O$_{4}$, and CuCr$_{2}$O$_{4}$  compounds while the magnetic transition in ZnCr$_{2}$O$_{4}$ is of first order.

\section{Conclusions}

In this work we study the magnetocaloric (MCE) response ($-\Delta S _M$) of all the spinels MnCr$_{2}$O$_{4}$,  FeCr$_{2}$O$_{4}$,  CoCr$_{2}$O$_{4}$, NiCr$_{2}$O$_{4}$, CuCr$_{2}$O$_{4}$, and ZnCr$_{2}$O$_{4}$ compounds across their magnetic transition $T_{C}$ and $T_{N}$.
The spinels ACr$_{2}$O$_{4}$ where A$^{2+}$ is a magnetic ion show paramagnetic to ferrimagnetic phase transition at temperature $T_{C}$. The spinels ACr$_{2}$O$_{4}$ where A$^{2+}$ is non-magnetic ion show paramagnetic to antiferromagnetic phase transition at temperature $T_{N}$. We observed that MnCr$_{2}$O$_{4}$ shows a maximum value of MCE in all the studied here spinel chromites while ZnCr$_{2}$O$_{4}$ shows inverse MCE. It is clear to say that the degree of MCE (-$\Delta S_M^{max}$) depends on the magnetic moment of A-site ion in ACr$_2$O$_4$. The MCE (-$\Delta S_M^{max}$) of the spinels decrease on decrease of A-site spin magnetic moment for all the spinels except for CuCr$_2$O$_4$. Further microscopic probe studies will be required to clarify the origin of this exception for CuCr$_2$O$_4$.  
       
\section{Acknowledgment}
We acknowledge the support of the X-ray facility at IISER Mohali for powder XRD measurements.

\begin{table*}[!htbp]
	\begin{center}
		\caption{Synthesis temperature conditions for ACr$_{2}$O$_{4}$.}
		\label{tab_spinel}
		\setlength\extrarowheight{4pt}
		\setlength{\tabcolsep}{14pt}
		\begin{tabular}{|c|c|c|c|c|c|c|}
			\hline
			\multirow{2}{*}{\textbf{ACr$_{2}$O$_{4}$}} & \multicolumn{3}{c|}{\textbf{Heat Treatments}} & \multirow{2}{*}{\textbf{Environment}}  \\
			\cline{2-4} & \textbf{First} & \textbf{Second} & \textbf{Third} &  \\ \hline
			NiCr$_{2}$O$_{4}$ 	& 800$^{\circ}$/24h & 1100$^{\circ}$/24h & 110$^{\circ}$/24h & in air  \\ \hline
			CoCr$_{2}$O$_{4}$ 	& 800$^{\circ}$/24h & 1000$^{\circ}$/24h & 1000$^{\circ}$/24h & in air  \\ \hline
			CuCr$_{2}$O$_{4}$ 	& 1100$^{\circ}$/24h & 1200$^{\circ}$/24h & 1200$^{\circ}$/24h & in air  \\ \hline
			ZnCr$_{2}$O$_{4}$ 	& 800$^{\circ}$/24h & 1200$^{\circ}$/24h & 1200$^{\circ}$/24h & in air  \\ \hline
			MnCr$_{2}$O$_{4}$ 	& 1100$^{\circ}$/24h & 1200$^{\circ}$/24h & 1300$^{\circ}$/24h & in inert   \\ \hline
			FeCr$_{2}$O$_{4}$ 	& 800$^{\circ}$/24h & 1000$^{\circ}$/24h & 13000$^{\circ}$/24h & in inert  \\ \hline
		\end{tabular}
	\end{center}
\end{table*}

\begin{table*}[htbp]
	\begin{center}
		\caption{Space group and refined lattice parameters of ACr$_2$O$_4$ for the powder X-ray diffraction data taken at room temperature.}
		\vspace*{0.5cm}
		\label{6_XRD}
		\setlength\extrarowheight{12pt}
		\setlength{\tabcolsep}{16pt}
		\begin{tabular}{|c|c|c|c|c|c|}
			\hline
			\hline
			\textbf{ACr$_2$O$_4$} & \textbf{Space group} & \textbf{a (A$^{\circ}$)} & \textbf{b (A$^{\circ}$)} & \textbf{c (A$^{\circ}$)} & $\mathbf{\alpha=\beta=\gamma}$ \\ \hline
			MnCr$_2$O$_4$ & Fd$\bar{3}$m & 8.4374(1) & 8.4374(1) & 8.4374(1) & 90$^{\circ}$ \\ \hline
			FeCr$_2$O$_4$ & Fd$\bar{3}$m & 8.3922(1) & 8.3922(1) & 8.3922(1) & 90$^{\circ}$ \\ \hline
			CoCr$_2$O$_4$ & Fd$\bar{3}$m & 8.3364(1) & 8.3364(1) & 8.3364(1) & 90$^{\circ}$ \\ \hline
			NiCr$_2$O$_4$ & I4$_1$/amd & 5.8763(1) & 5.8763(1) & 8.3180(2) & 90$^{\circ}$ \\ \hline
			CuCr$_2$O$_4$ & I4$_1$/amd & 6.0327(2) & 6.0327(2) & 7.7958(3) & 90$^{\circ}$ \\ \hline
			ZnCr$_2$O$_4$ & Fd$\bar{3}$m & 8.3284(1) & 8.3284(1) & 8.3284(1) & 90$^{\circ}$ \\ \hline
			\hline 
		\end{tabular}
	\end{center}
\end{table*}

\begin{table*}[htbp]
	
	\begin{center}
		\caption{Table for each A site ion the value of S, effective magnetic moment \cite{Blundell}, magnetocaloric parameters and transition temperature for all the ACr$_{2}$O$_{4}$ where A = Mn, Fe, Co, Ni, Cu and Zn compounds in a external magnetic field of strength 9T).}
		\vspace*{0.5cm}
		\label{tab}
		\setlength\extrarowheight{12pt}
		\setlength{\tabcolsep}{12pt}
		
		\begin{tabular}{|c|c|c|c|c|c|c|}
			\hline
			\hline
			\textbf{ACr$_2$O$_4$}  & \textbf{S} & \textbf{M($\mu_{B}$)} & \textbf{-$\Delta$S$_{M}$(J/kg-K)} & \textbf{$\Delta$T$_{FWHM}$(K)} & \textbf{RCP (J/kg)}    & \textbf{T$_{C}$} or \textbf{T$_{N}$(K)} \\ \hline
			MnCr$_2$O$_4$  & 2.5       & 5.92     & 7.16   & 28     & 200.48   & 41       \\ \hline
			FeCr$_2$O$_4$  & 2.0       & 4.90     & 2.35   & 28     & 65.8     & 101       \\ \hline
			CoCr$_2$O$_4$  & 1.5       & 3.87     & 2.25   & 17   & 38.35     & 97       \\ \hline
			NiCr$_2$O$_4$  & 1.0       & 2.83     & 1.5    & 25     & 37.5     & 68       \\ \hline
			CuCr$_2$O$_4$  & 0.5       & 1.73     & 3.22   & 24     & 77.28    & 129       \\ \hline
			ZnCr$_2$O$_4$  & 0.0       & 0.0      & -0.2   & 2.34   & 0.468    & 13       \\ \hline
			CdCr$_2$S$_4$ \cite{doi:10.1063/1.2751576}   & --        & --       & 7.04 at 4T &  -- & -- & 87 \\ \hline
			MnV$_2$O$_4$ \cite{Luo_2009}   & --        & --       & 24.0 at 4T & -- & -- & 57 \\ \hline
			Cd$_{0.8}$Cu$_{0.2}$Cr$_2$S$_4$ \cite{doi:10.1063/1.2830973}   & --        & --       & 5.1 at 5T & -- & -- & 86 \\ \hline     
			\hline
		\end{tabular}
	\end{center}
\end{table*}

\bibliographystyle{apsrev4-1}
\bibliography{Ref}
\nocite{*}
\end{document}